\documentclass[sigconf]{acmart}

\usepackage{ctable}
\usepackage{multirow}
\usepackage{balance}
\usepackage{booktabs}
\usepackage{hyphsubst}
\usepackage{graphicx}
\usepackage{amsmath}
\usepackage{url}
\usepackage{breakurl}
\usepackage{hhline}
\usepackage[caption = false]{subfig}

\newcounter{lizcounter}
\DeclareRobustCommand{\liz}[1]{\textbf{/* #1 (liz) */}\stepcounter{lizcounter}\typeout{LaTeX Warning: liz comment \thelizcounter: #1 (line \the\inputlineno)}}

\newcounter{findingscounter}

\newcommand{\para}[1]{\vspace{2mm}\noindent\textbf{#1}}

\copyrightyear{2018} 
\acmYear{2018} 
\setcopyright{acmcopyright}
\acmConference{SIR@CIKM'2018}{October 22-26, 2017}{Turin, Italy}\acmPrice{15.00}\acmDOI{xxx}
\acmISBN{xxx}

\begin{document}

\title{AFEL-REC: A Recommender System for Providing Learning Resource Recommendations in Social Learning Environments}


\author{Dominik Kowald, Emanuel Lacic, Dieter Theiler \& Elisabeth Lex}
\affiliation{
  \institution{Know-Center GmbH \& Graz University of Technology, Austria}
}
\email{{dkowald,elacic,dtheiler}@know-center.at,elisabeth.lex@tugraz.at}

\begin{abstract}
In this paper, we present preliminary results of AFEL-REC, a recommender system for social learning environments. AFEL-REC is build upon a scalable software architecture to provide recommendations of learning resources in near real-time. Furthermore, AFEL-REC can cope with any kind of data that is present in social learning environments such as resource metadata, user interactions or social tags. We provide a preliminary evaluation of three recommendation use cases implemented in AFEL-REC and we find that utilizing social data in form of tags is helpful for not only improving recommendation accuracy but also coverage. This paper should be valuable for both researchers and practitioners interested in providing resource recommendations in social learning environments.
\end{abstract}


\maketitle

\section{Introduction}
Recommender systems aim to predict if a specific user will like a specific resource. To do so, recommender systems analyze past usage behavior (e.g., clicks or likes) with the goal to generate a personalized list of potentially relevant resources \cite{ricci2011introduction}. Nowadays, recommender systems are part of many applications, such as online marketplaces (e.g., Amazon), movie streaming services (e.g., Netflix), job portals (e.g., LinkedIn), and Technology Enhanced Learning (TEL) environments (e.g., Coursera).

Especially in the field of TEL, recommender systems have become an important research area over the past decade \cite{drachsler2015panorama}. One of the many examples in this area is the CoFind system \cite{dron2000cofind}, which guides learners to resources that were found useful by other learners in the past. Other examples include the work of \cite{mangina2008evaluation}, who proposed a recommendation approach with query extraction mechanisms or the work of \cite{gomez2009recommendation}, who enhanced Collaborative Filtering (CF) \cite{herlocker2004evaluating} by taking into account the learner's evolution in time. Another recent strand is the research on context-aware recommender approaches for TEL. Here, contextual information, such as the location \cite{yu2010towards}, are incorporated into the recommendation process. In this respect, social learning environments \cite{vassileva2008toward}, which aim to support users in learning through the observation of other users' behaviors, bear great potential for recommender systems as they provide a vast amount of social information. This includes, for example, friendship connections, group memberships or social tags, which are freely-chosen keywords used for collaboratively annotating learning resources \cite{kowald2015refining}. Although a lot of recommender systems and algorithms are available in the TEL area, there is still the lack of research on recommender systems specifically tailored for social learning environments (such as e.g., \cite{el20103a}).

Therefore, in the course of the H2020 project Analytics for Everyday Learning (AFEL)\footnote{\url{http://afel-project.eu/}}, we have developed AFEL-REC, which is a recommender system for social learning environments. AFEL-REC is build upon a scalable software architecture to support various use cases for providing recommendations of learning resources in near real-time (see Section \ref{s:approach}). We conducted a preliminary evaluation of AFEL-REC using data gathered from the Spanish social learning environment Didactalia\footnote{\url{https://didactalia.net/}} (see Section \ref{s:eval}). Taken together, our contributions are twofold (see also Section \ref{s:conc}):
\begin{enumerate}
\item We present AFEL-REC and use cases for providing resource recommendations in social learning environments.
\item We demonstrate that social information, such as social tags, can be used to improve the accuracy and coverage of recommendations in social learning environments.
\end{enumerate}
We believe that our work contributes to the under-researched portfolio of recommender systems for social learning environments. Furthermore, we present an overview of use cases and preliminary evaluation results, which should be valuable for both researchers and practitioners interested in providing resource recommendations in social learning environments.

\section{Approach} \label{s:approach}
In this section, we present our AFEL-REC system by providing a detailed description of its software architecture as well as potential use cases that can be realized for recommending resources in social learning environments.

\subsection{System Overview} \label{s:framework}
The software architecture of AFEL-REC is based on the scalable recommendation framework ScaR\footnote{\url{http://scar.know-center.tugraz.at/}} \cite{lacic2015scar}. It is depicted in Figure \ref{fig:afel} and consists of the following main modules:

\begin{figure*}[t!]
   \centering
	\includegraphics[width=0.7\textwidth]{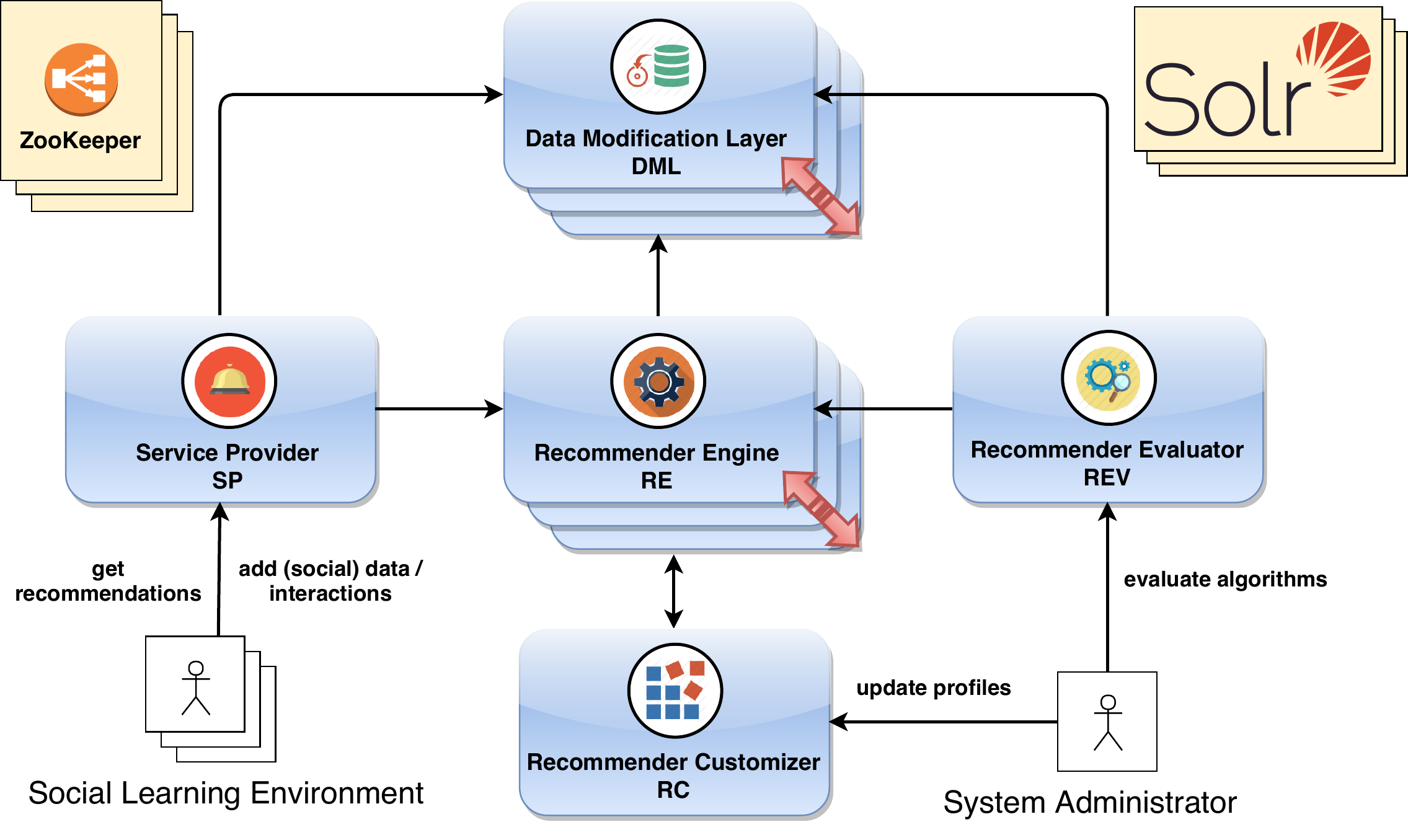}
   \caption{The scalable AFEL-REC software architecture based on the powerful open-source frameworks Apache Solr and Apache ZooKeeper. The main modules of AFEL-REC communicate via REST-based Web services.\vspace{-3mm}}
	 \label{fig:afel}
\end{figure*}

\para{Service Provider (SP).} The SP acts as a proxy for social learning environments to access AFEL-REC. Thus, it provides REST-based Web services to enable clients to query recommendations and to add new data (e.g., user interactions or learning resources) to the recommender system.

\para{Data Modification Layer (DML) \& Solr.} The DML encapsulates all CRUD operations (i.e., create, retrieve, update, delete) in one module and therefore, enables easy access to the underlying data backend. As depicted in Figure \ref{fig:afel}, AFEL-REC uses the high-performance search platform Apache Solr\footnote{\url{http://lucene.apache.org/solr/}}. This data backend solution not only guarantees scalability and near real-time recommendations but also the support of multiple data sources. While most recommender systems rely on rating-based data, AFEL-REC is also capable of processing relevant social information such as tags.

\para{Recommender Engine (RE).} The RE is the heart of AFEL-REC and is responsible for calculating recommendations. As we are using Apache Solr, we can benefit from its build-in data structures for efficiently calculating user and resource similarities, and ranking resources based on their relevance for a specific recommendation context. In Section \ref{s:use}, we discuss possible use cases and algorithms that can be realized using AFEL-REC's recommender engine.

\para{Recommender Customizer (RC).} The RC is used to change the parameters (e.g., the neighborhood size $n$) of the recommendation approaches on the fly. Thus, it holds a so-called recommendation profile for each approach, which can be accessed and changed by the system administrator. These changes are then broadcast to the RE to be aware of how a specific approach should be executed.

\para{Recommender Evaluator (REV).} The REV is responsible for evaluating the recommendation algorithms implemented in the RE. Thus, it can be executed to perform an offline evaluation with training/test set splits (see Section \ref{s:eval}) or an online evaluation with A/B-tests.

\para{ZooKeeper.} We are using Apache ZooKeeper\footnote{\url{https://zookeeper.apache.org/}} for handling the communication between the modules and for offering horizontal scaling. Thus, in cases in which we observe a high request load, we can start multiple instances of the same module (indicated by the arrows in the DML and RE modules).

\subsection{Use Cases} \label{s:use}
Using this software architecture, AFEL-REC is capable of supporting seven use cases for providing recommendations in social learning environments:

\para{UC1: Recommendation of Popular Resources in the Social Learning Environment.}
The first use case is a non-personalized one and is especially useful for new users of a social learning environment without any user interactions so far (i.e., cold-start users  \cite{schein2002methods}). Thus, it is typically realized using a MostPopular algorithm. This approach recommends learning resources, which are weighted and ranked by the number of interactions. As mentioned, the MostPopular approach is non-personalized and thus, each user will receive the same recommendations.

\para{UC2: Recommendation of Resources That Like-Minded Users Have Interacted With.}
The second use case is a personalized one and thus, analyzes past user interactions to specifically tailor recommendations towards learners. Collaborative Filtering (CF) algorithms are typically chosen to realize such a use case \cite{herlocker2004evaluating}. CF approaches analyze the interactions between users and items (e.g., learning resources) and recommend those items to a given user that similar users have interacted with in the past. More specifically, in CF methods two users are treated as similar if they have liked the same items in the past. This in turn allows us to assume that these two users will also like the same (or similar) items in the future.

\para{UC3: Recommendation of Resources Based on Social Information.}
This use case is similar to UC2 but this time two users are treated as similar if they have shared some social information in the past. This social information can be friendship connections, group memberships or social tags. We hypothesize that social information is capable of providing a richer semantic representation of a user's interests than pure interaction data. Therefore, we also think that this should positively influence the coverage of the recommendations as validated in Section \ref{s:eval} of this paper.

\para{UC4: Recommendation of Resources That are Similar to the Resources the User Has Interacted With.}
One disadvantage of UC2 and UC3 is that it can only be applied to resources, which already have user interactions or social information attached to them. This means that cold-start resources without any user interactions or social information cannot be recommended. To overcome this, UC3 aims at utilizing resource similarities for personalized recommendations by using a Content-based Filtering (CBF) approach \cite{lops2011content}. CBF methods use resource features such as categories or description texts for calculating similarities between resources. Then, the most similar resources of the resources the given user has interacted with are recommended. 

\para{UC5: Recommendation of (Alternative) Resources for a Specific Resource.}
This use case is related to UC4 but provides contextualized recommendations instead of personalized ones. This means that recommendations are not based on the learner but based on a specific resource by finding alternative ones. Thus, similar to UC4, the most similar resources for the given resource are recommended using CBF.

\para{UC6: Recommendation of (Alternative) Resources for a Specific User and a Specific Resource.}
The next use case also focuses on recommending alternative resources for a specific resource but this time in a personalized manner. Such use case can be implemented using a contextualized CF approach. This means that we search for similar users of the target user, who have also interacted with the target resource. Thus, this use case is similar to Amazon's ``Users who bought this, also bought that'' recommender.

\para{UC7: Recommendation of Resources for a Specific User and a Specific Learning Goal.} 
Finally, the last use case tackles recommendations in a personalized and adaptive manner. While UC1 to UC6 have focused on providing relevant recommendations, they have neglected the learning goal of the user. Such a learning goal could be the aim to focus on a specific topic or to receive more difficult learning resources. To realize adaptive recommendations, the suggested resources could be re-ranked using a feature boosting technique. For example, if the learning goal is to study more difficult resources, then resources with a higher complexity (e.g., measured via a readability score) should be boosted and easy ones should be down-graded in the recommendation list.

\section{Preliminary Evaluation} \label{s:eval}
In this section, we present preliminary evaluation results for AFEL-REC. The aim of this evaluation is twofold: (i) we want to show that AFEL-REC is capable of providing recommendations using data gathered from a real-world social learning environment, and (ii), we want show that social information can enhance the prediction accuracy and coverage of recommendations. To do so, we focus on UC1 to UC3 presented in the previous section.

\subsection{Data}
We collected our data from the social learning environment Didactalia between the 26th of February 2017 until the 28th of May 2018. This included 1,879,761 user interactions (i.e., clicks on learning resources) by 1,274,858 users on 35,346 learning resources. This resulted in 1.47 interactions per user and 53.18 interactions per learning resource on average. Additionally, 485,295 social tags were applied to these learning resources, which resulted in 13.73 tags per resource on average. The full statistics of our dataset are shown in Table \ref{tab:datasets}. To date, the only social information we are using in our data are tags but we are planning to extend this by also incorporating connections between users or group memberships \cite{lacic2015utilizing}.

\begin{table}[t!]
	\small
  \setlength{\tabcolsep}{2.0pt}	
  \centering	
    \begin{tabular}{l|c}
    \specialrule{.2em}{.1em}{.1em}
											Number of interactions (i.e., clicks)										& 1,879,761			\\
											Number of users																					&	1,274,858			\\
											Number of learning resources														& 35,346  			\\
											Number of social tags	  																& 485,295  			\\\hline
											Average number of interactions per user									& 1.47					\\
											Average number of interactions per learning resource 		& 53.18					\\
											Average number of tags per learning resource						& 13.73			 		\\
		\specialrule{.2em}{.1em}{.1em}								
    \end{tabular}
	  \caption{Statistics of our dataset, which was collected from the social learning environment Didactalia.\vspace{-6mm}}
  \label{tab:datasets}
\end{table}

\subsection{Evaluation Method and Metrics}
For evaluating AFEL-REC, we split our dataset into training and test sets. Therefore, we followed common practice in the research area of recommender systems and information retrieval by using the most recent 20\% of interactions of each user for testing and the remaining 80\% for training. This dataset splitting technique ensures that the chronological order of the data is preserved and thus, that the future is predicted based on past user interactions.

For measuring the accuracy of the recommendations, we use a rich set of metrics, namely Recall (R@20, measured for $k$ = 20 recommended resources), Precision (P@1, for $k$ = 1), F1-score (F1@10 for $k$ = 10), Mean Reciprocal Rank (MRR@20, for $k$ = 20), Mean Average Precision (MAP@20, for $k$ = 20) and normalized Discounted Cumulative Gain (nDCG@20, for $k$ = 20) \cite{trattner2015tagrec}. Furthermore, we also report the coverage (C) of the recommendations, measuring the fraction of users for whom the algorithm is capable of producing any recommendations.

\subsection{Recommendation Approaches}
We evaluated UC1 - UC3 presented in Section \ref{s:use} to show (i) the general usefulness of AFEL-REC for providing recommendations in social learning environments, and (ii) that social information in the form of tags is helpful for improving the recommendation accuracy and coverage. In the future, we will also evaluate UC4 - UC7.

The MP (MostPopular) approach refers to UC1 and is a non-personalized algorithm, which recommends the most frequently used learning resources in the system. This algorithm also works for cold-start users \cite{schein2002methods} and thus, should reach a UC of 100\%.

The CF$_i$ algorithm refers to UC2 and calculates the neighborhood of a user on the basis of interaction data (i.e., clicks). Thus, two users are treated as similar if they have interacted with the same learning resources in the past \cite{herlocker2004evaluating}. Based on \cite{kopeinik2017improving}, we used a neighborhood size of $n$ = 20 users.

Similar to CF$_i$, CF$_t$ is a Collaborative Filtering-based approach but, as discussed in UC3, this one calculates the user neighborhood based on social tags. Thus, two users are treated as similar if they have used the same social tags in the past \cite{parra2009collaborative}.

\begin{table}[t!]
	\small
  \setlength{\tabcolsep}{2.0pt}	
  \centering
    \begin{tabular}{c|cccccc|c}
    \specialrule{.2em}{.1em}{.1em}
											Approach				& R@20								& P@1						& F1@10					& MRR@20					& MAP@20					& nDCG@20			& C 							\\\hline
											UC1: MP					& .007						& .002					& .002					& .002						& .002						& .003						& \textbf{100\%}	\\
											UC2: CF$_i$			&	.046						& .022 					& .012 					& .025						& .026						& .032						& 40\%						\\
											UC3: CF$_t$			& \textbf{.070}	  & \textbf{.027}	& \textbf{.016}	& \textbf{.034}		& \textbf{.035}		& \textbf{.044}		& 53\%						\\
		\specialrule{.2em}{.1em}{.1em}								
    \end{tabular}
    \caption{Preliminary results of our evaluation of AFEL-REC. We see that CF$_t$ provides not only a better recommendation accuracy but also coverage (C) than CF$_i$, which means that social information in the form of tags is helpful for improving recommendations in social learning environments.\vspace{-6mm}}		
  \label{tab:results}
\end{table}

\subsection{Results}
The preliminary results of our evaluation are shown in Table \ref{tab:results}. While the lowest accuracy estimates are provided by the unpersonalized MP approach (UC1), this approach also provides the highest UC with 100\%. This means that MP is capable of providing recommendations for all users in the social learning environment.

When looking at the next algorithm, CF$_i$ (i.e., CF on the basis of user interactions, UC2), we notice that this approach provides approx. 10 times higher accuracy values than MP. We contribute this to the personalization factor of the algorithm. However, one drawback of CF$_i$ is the rather small UC of 40\%, which means that it cannot generate any recommendations for 60\% of the users.

Finally, CF$_t$ (i.e., CF on the basis of social tags, UC3), provides not only the best results with respect to recommendation accuracy but also a larger coverage (C) than CF$_i$. This result shows that social information can indeed help to improve recommendations in social learning environments.

\section{Conclusion and Future Work} \label{s:conc}
In this paper, we presented AFEL-REC, a recommender system for social learning environments. AFEL-REC is build upon a scalable software architecture based on powerful open-source frameworks such as Apache Solr or Apache ZooKeeper and thus, is capable of providing near real-time recommendations of learning resources. We have demonstrated the usefulness of AFEL-REC by discussing use cases that can be realized with it and by providing preliminary evaluation results using data gathered from the Spanish social learning environment Didactalia. Our evaluation results show that social information in the form of tags can be used to enhance the accuracy and coverage of learning resource recommendations. These findings should be of interest for both researchers and practitioners interested in providing resource recommendations in social learning environments.

One limitation of this work is that we have only investigated tags as potential social information for learning resource recommendations. Thus, for future work, we plan to extend AFEL-REC by also considering additional types of social information such as friendship connections or group memberships. Also, we plan to include our cognitive-inspired tag recommender algorithms, which have been specifically useful in the context of TEL \cite{Kowald2017,Kopeinik2017}. Finally, we will also evaluate the remaining five use cases, which we have not tackled so far.

\para{Acknowledgments.} The authors would like to thank Didactalia and the AFEL consortium. This work was supported by the Know-Center GmbH Graz (Austrian FFG COMET Program) and the European-funded H2020 project AFEL (GA: 687916).

\bibliographystyle{ACM-Reference-Format}

\begin{thebibliography}{00}


\ifx \showCODEN    \undefined \def \showCODEN     #1{\unskip}     \fi
\ifx \showDOI      \undefined \def \showDOI       #1{{\tt DOI:}\penalty0{#1}\ }
  \fi
\ifx \showISBNx    \undefined \def \showISBNx     #1{\unskip}     \fi
\ifx \showISBNxiii \undefined \def \showISBNxiii  #1{\unskip}     \fi
\ifx \showISSN     \undefined \def \showISSN      #1{\unskip}     \fi
\ifx \showLCCN     \undefined \def \showLCCN      #1{\unskip}     \fi
\ifx \shownote     \undefined \def \shownote      #1{#1}          \fi
\ifx \showarticletitle \undefined \def \showarticletitle #1{#1}   \fi
\ifx \showURL      \undefined \def \showURL       #1{#1}          \fi
\providecommand\bibfield[2]{#2}
\providecommand\bibinfo[2]{#2}
\providecommand\natexlab[1]{#1}
\providecommand\showeprint[2][]{arXiv:#2}

\bibitem[\protect\citeauthoryear{Drachsler, Verbert, Santos, and
  Manouselis}{Drachsler et~al\mbox{.}}{2015}]%
        {drachsler2015panorama}
\bibfield{author}{\bibinfo{person}{Hendrik Drachsler}, \bibinfo{person}{Katrien
  Verbert}, \bibinfo{person}{Olga~C Santos}, {and} \bibinfo{person}{Nikos
  Manouselis}.} \bibinfo{year}{2015}\natexlab{}.
\newblock \showarticletitle{Panorama of recommender systems to support
  learning}.
\newblock In \bibinfo{booktitle}{{\em Recommender systems handbook}}.
  \bibinfo{publisher}{Springer}, \bibinfo{pages}{421--451}.
\newblock


\bibitem[\protect\citeauthoryear{Dron, Mitchell, Siviter, and Boyne}{Dron
  et~al\mbox{.}}{2000}]%
        {dron2000cofind}
\bibfield{author}{\bibinfo{person}{Jon Dron}, \bibinfo{person}{Richard
  Mitchell}, \bibinfo{person}{Phil Siviter}, {and} \bibinfo{person}{Chris
  Boyne}.} \bibinfo{year}{2000}\natexlab{}.
\newblock \showarticletitle{CoFIND-an experiment in n-dimensional collaborative
  filtering}.
\newblock \bibinfo{journal}{{\em Journal of Network and Computer
  Applications\/}} \bibinfo{volume}{23}, \bibinfo{number}{2}
  (\bibinfo{year}{2000}), \bibinfo{pages}{131--142}.
\newblock


\bibitem[\protect\citeauthoryear{El~Helou, Salzmann, and Gillet}{El~Helou
  et~al\mbox{.}}{2010}]%
        {el20103a}
\bibfield{author}{\bibinfo{person}{Sandy El~Helou}, \bibinfo{person}{Christophe
  Salzmann}, {and} \bibinfo{person}{Denis Gillet}.}
  \bibinfo{year}{2010}\natexlab{}.
\newblock \showarticletitle{The 3A Personalized, Contextual and Relation-based
  Recommender System.}
\newblock \bibinfo{journal}{{\em J. UCS\/}}  \bibinfo{volume}{16}
  (\bibinfo{year}{2010}).
\newblock


\bibitem[\protect\citeauthoryear{Gomez-Albarran and
  Jimenez-Diaz}{Gomez-Albarran and Jimenez-Diaz}{2009}]%
        {gomez2009recommendation}
\bibfield{author}{\bibinfo{person}{Mercedes Gomez-Albarran} {and}
  \bibinfo{person}{Guillermo Jimenez-Diaz}.} \bibinfo{year}{2009}\natexlab{}.
\newblock \showarticletitle{Recommendation and students’ authoring in
  repositories of learning objects: A case-based reasoning approach}.
\newblock \bibinfo{journal}{{\em International Journal of Emerging Technologies
  in Learning (iJET)\/}} \bibinfo{volume}{4}, \bibinfo{number}{2009}
  (\bibinfo{year}{2009}), \bibinfo{pages}{35--40}.
\newblock


\bibitem[\protect\citeauthoryear{Herlocker, Konstan, Terveen, and
  Riedl}{Herlocker et~al\mbox{.}}{2004}]%
        {herlocker2004evaluating}
\bibfield{author}{\bibinfo{person}{Jonathan~L Herlocker},
  \bibinfo{person}{Joseph~A Konstan}, \bibinfo{person}{Loren~G Terveen}, {and}
  \bibinfo{person}{John~T Riedl}.} \bibinfo{year}{2004}\natexlab{}.
\newblock \showarticletitle{Evaluating collaborative filtering recommender
  systems}.
\newblock \bibinfo{journal}{{\em ACM Transactions on Information Systems
  (TOIS)\/}} \bibinfo{volume}{22}, \bibinfo{number}{1} (\bibinfo{year}{2004}),
  \bibinfo{pages}{5--53}.
\newblock


\bibitem[\protect\citeauthoryear{Kopeinik, Kowald, Hasani-Mavriqi, and
  Lex}{Kopeinik et~al\mbox{.}}{2017a}]%
        {kopeinik2017improving}
\bibfield{author}{\bibinfo{person}{Simone Kopeinik}, \bibinfo{person}{Dominik
  Kowald}, \bibinfo{person}{Ilire Hasani-Mavriqi}, {and}
  \bibinfo{person}{Elisabeth Lex}.} \bibinfo{year}{2017}\natexlab{a}.
\newblock \showarticletitle{Improving Collaborative Filtering Using a Cognitive
  Model of Human Category Learning}.
\newblock \bibinfo{journal}{{\em The Journal of Web Science\/}}
  \bibinfo{volume}{2}, \bibinfo{number}{1} (\bibinfo{year}{2017}).
\newblock


\bibitem[\protect\citeauthoryear{Kopeinik, Lex, Seitlinger, Albert, and
  Ley}{Kopeinik et~al\mbox{.}}{2017b}]%
        {Kopeinik2017}
\bibfield{author}{\bibinfo{person}{Simone Kopeinik}, \bibinfo{person}{Elisabeth
  Lex}, \bibinfo{person}{Paul Seitlinger}, \bibinfo{person}{Dietrich Albert},
  {and} \bibinfo{person}{Tobias Ley}.} \bibinfo{year}{2017}\natexlab{b}.
\newblock \showarticletitle{Supporting Collaborative Learning with Tag
  Recommendations: A Real-world Study in an Inquiry-based Classroom Project}.
  In \bibinfo{booktitle}{{\em Proceedings of the Seventh International Learning
  Analytics and Knowledge Conference}} {\em (\bibinfo{series}{LAK '17})}.
  \bibinfo{publisher}{ACM}, \bibinfo{address}{New York, NY, USA},
  \bibinfo{pages}{409--418}.
\newblock
\showISBNx{978-1-4503-4870-6}
\showDOI{%
\url{http://dx.doi.org/10.1145/3027385.3027421}}


\bibitem[\protect\citeauthoryear{Kowald, Kopeinik, Seitlinger, Ley, Albert, and
  Trattner}{Kowald et~al\mbox{.}}{2015}]%
        {kowald2015refining}
\bibfield{author}{\bibinfo{person}{Dominik Kowald}, \bibinfo{person}{Simone
  Kopeinik}, \bibinfo{person}{Paul Seitlinger}, \bibinfo{person}{Tobias Ley},
  \bibinfo{person}{Dietrich Albert}, {and} \bibinfo{person}{Christoph
  Trattner}.} \bibinfo{year}{2015}\natexlab{}.
\newblock \showarticletitle{Refining frequency-based tag reuse predictions by
  means of time and semantic context}.
\newblock In \bibinfo{booktitle}{{\em Mining, Modeling, and
  Recommending'Things' in Social Media}}. \bibinfo{publisher}{Springer},
  \bibinfo{pages}{55--74}.
\newblock


\bibitem[\protect\citeauthoryear{Kowald, Pujari, and Lex}{Kowald
  et~al\mbox{.}}{2017}]%
        {Kowald2017}
\bibfield{author}{\bibinfo{person}{Dominik Kowald},
  \bibinfo{person}{Subhash~Chandra Pujari}, {and} \bibinfo{person}{Elisabeth
  Lex}.} \bibinfo{year}{2017}\natexlab{}.
\newblock \showarticletitle{Temporal Effects on Hashtag Reuse in Twitter: A
  Cognitive-Inspired Hashtag Recommendation Approach}. In
  \bibinfo{booktitle}{{\em Proceedings of the 26th International Conference on
  World Wide Web}} {\em (\bibinfo{series}{WWW '17})}.
  \bibinfo{pages}{1401--1410}.
\newblock
\showISBNx{978-1-4503-4913-0}


\bibitem[\protect\citeauthoryear{Lacic, Kowald, Eberhard, Trattner, Parra, and
  Marinho}{Lacic et~al\mbox{.}}{2015a}]%
        {lacic2015utilizing}
\bibfield{author}{\bibinfo{person}{Emanuel Lacic}, \bibinfo{person}{Dominik
  Kowald}, \bibinfo{person}{Lukas Eberhard}, \bibinfo{person}{Christoph
  Trattner}, \bibinfo{person}{Denis Parra}, {and}
  \bibinfo{person}{Leandro~Balby Marinho}.} \bibinfo{year}{2015}\natexlab{a}.
\newblock \showarticletitle{Utilizing online social network and location-based
  data to recommend products and categories in online marketplaces}.
\newblock In \bibinfo{booktitle}{{\em Mining, Modeling, and
  Recommending'Things' in Social Media}}. \bibinfo{publisher}{Springer}.
\newblock


\bibitem[\protect\citeauthoryear{Lacic, Traub, Kowald, and Lex}{Lacic
  et~al\mbox{.}}{2015b}]%
        {lacic2015scar}
\bibfield{author}{\bibinfo{person}{Emanuel Lacic}, \bibinfo{person}{Matthias
  Traub}, \bibinfo{person}{Dominik Kowald}, {and} \bibinfo{person}{Elisabeth
  Lex}.} \bibinfo{year}{2015}\natexlab{b}.
\newblock \showarticletitle{ScaR: Towards a Real-Time Recommender Framework
  Following the Microservices Architecture}. In \bibinfo{booktitle}{{\em
  Proceedings of LSRS2015 Workshop at RecSys 2015}}. .
\newblock


\bibitem[\protect\citeauthoryear{Lops, De~Gemmis, and Semeraro}{Lops
  et~al\mbox{.}}{2011}]%
        {lops2011content}
\bibfield{author}{\bibinfo{person}{Pasquale Lops}, \bibinfo{person}{Marco
  De~Gemmis}, {and} \bibinfo{person}{Giovanni Semeraro}.}
  \bibinfo{year}{2011}\natexlab{}.
\newblock \showarticletitle{Content-based recommender systems: State of the art
  and trends}.
\newblock In \bibinfo{booktitle}{{\em Recommender systems handbook}}.
  \bibinfo{publisher}{Springer}, \bibinfo{pages}{73--105}.
\newblock


\bibitem[\protect\citeauthoryear{Mangina and Kilbride}{Mangina and
  Kilbride}{2008}]%
        {mangina2008evaluation}
\bibfield{author}{\bibinfo{person}{Eleni Mangina} {and} \bibinfo{person}{John
  Kilbride}.} \bibinfo{year}{2008}\natexlab{}.
\newblock \showarticletitle{Evaluation of keyphrase extraction algorithm and
  tiling process for a document/resource recommender within e-learning
  environments}.
\newblock \bibinfo{journal}{{\em Computers \& Education\/}}
  \bibinfo{volume}{50}, \bibinfo{number}{3} (\bibinfo{year}{2008}),
  \bibinfo{pages}{807--820}.
\newblock


\bibitem[\protect\citeauthoryear{Parra and Brusilovsky}{Parra and
  Brusilovsky}{2009}]%
        {parra2009collaborative}
\bibfield{author}{\bibinfo{person}{Denis Parra} {and} \bibinfo{person}{Peter
  Brusilovsky}.} \bibinfo{year}{2009}\natexlab{}.
\newblock \showarticletitle{Collaborative filtering for social tagging systems:
  an experiment with CiteULike}. In \bibinfo{booktitle}{{\em Proceedings of the
  third ACM conference on Recommender systems}}. ACM,
  \bibinfo{pages}{237--240}.
\newblock


\bibitem[\protect\citeauthoryear{Ricci, Rokach, and Shapira}{Ricci
  et~al\mbox{.}}{2011}]%
        {ricci2011introduction}
\bibfield{author}{\bibinfo{person}{Francesco Ricci}, \bibinfo{person}{Lior
  Rokach}, {and} \bibinfo{person}{Bracha Shapira}.}
  \bibinfo{year}{2011}\natexlab{}.
\newblock \bibinfo{booktitle}{{\em Introduction to recommender systems
  handbook}}.
\newblock \bibinfo{publisher}{Springer}.
\newblock


\bibitem[\protect\citeauthoryear{Schein, Popescul, Ungar, and Pennock}{Schein
  et~al\mbox{.}}{2002}]%
        {schein2002methods}
\bibfield{author}{\bibinfo{person}{Andrew~I Schein},
  \bibinfo{person}{Alexandrin Popescul}, \bibinfo{person}{Lyle~H Ungar}, {and}
  \bibinfo{person}{David~M Pennock}.} \bibinfo{year}{2002}\natexlab{}.
\newblock \showarticletitle{Methods and metrics for cold-start
  recommendations}. In \bibinfo{booktitle}{{\em Proceedings of the 25th annual
  international ACM SIGIR conference on Research and development in information
  retrieval}}. ACM, \bibinfo{pages}{253--260}.
\newblock


\bibitem[\protect\citeauthoryear{Trattner, Kowald, and Lacic}{Trattner
  et~al\mbox{.}}{2015}]%
        {trattner2015tagrec}
\bibfield{author}{\bibinfo{person}{Christoph Trattner},
  \bibinfo{person}{Dominik Kowald}, {and} \bibinfo{person}{Emanuel Lacic}.}
  \bibinfo{year}{2015}\natexlab{}.
\newblock \showarticletitle{TagRec: towards a toolkit for reproducible
  evaluation and development of tag-based recommender algorithms}.
\newblock \bibinfo{journal}{{\em ACM SIGWEB Newsletter\/}}
  \bibinfo{number}{Winter} (\bibinfo{year}{2015}), \bibinfo{pages}{3}.
\newblock


\bibitem[\protect\citeauthoryear{Vassileva}{Vassileva}{2008}]%
        {vassileva2008toward}
\bibfield{author}{\bibinfo{person}{Julita Vassileva}.}
  \bibinfo{year}{2008}\natexlab{}.
\newblock \showarticletitle{Toward social learning environments}.
\newblock \bibinfo{journal}{{\em IEEE transactions on learning technologies\/}}
  \bibinfo{volume}{1}, \bibinfo{number}{4} (\bibinfo{year}{2008}),
  \bibinfo{pages}{199--214}.
\newblock


\bibitem[\protect\citeauthoryear{Yu, Zhou, and Shu}{Yu et~al\mbox{.}}{2010}]%
        {yu2010towards}
\bibfield{author}{\bibinfo{person}{Zhiwen Yu}, \bibinfo{person}{Xingshe Zhou},
  {and} \bibinfo{person}{Lei Shu}.} \bibinfo{year}{2010}\natexlab{}.
\newblock \showarticletitle{Towards a semantic infrastructure for context-aware
  e-learning}.
\newblock \bibinfo{journal}{{\em Multimedia Tools and Applications\/}}
  \bibinfo{volume}{47}, \bibinfo{number}{1} (\bibinfo{year}{2010}).
\newblock


\end{thebibliography}

\end{document}